\documentclass[a4,11pt]{article}

\usepackage[latin1]{inputenc}

\usepackage{color}
\usepackage{graphicx}  

\usepackage{psfrag}

\usepackage{tocbibind} 

\usepackage{epsf}
\usepackage{epic}
\usepackage{eepic}

\usepackage{epsfig}
\usepackage{wrapfig}
\usepackage{amsmath,amssymb,amsfonts}
\usepackage[mathscr]{eucal}
\usepackage{citesort}
\usepackage{url}
\usepackage{fancybox}
\usepackage{ntheorem}

\newcommand{\hz}{{\hat z}{}}%

\newcommand{\ttheta}{\tilde\theta}

\newcommand{\ih}{instantaneous horizon}

\newcommand{\nih}{nondegenerate instantaneous horizon}

\newcommand{\uS}{U_{\mathrm{Schw}}}
\newcommand{\lS}{\lambda_{\mathrm{Schw}}}

\newcommand{\rS}{r_{\mathrm{Schw}}}

\newcommand{\zU}{\mathring U}
\newcommand{\zlambda}{\mathring \lambda}
\newcommand{\zbeta}{\mathring \beta}

\newcommand{\puncti}{a_i}

\newcommand{\fourg}{^4g}

\newcommand{\beijing}[1]{}

\newcommand{\tU}{\tilde U}

\newcommand{\MUone}{M/\mathrm{U(1)}}%
\newcommand{\Uone}{\mathrm{U(1)}}%

\newcommand{\Mextone}{\Mext/\mathrm{U(1)}}%

\newcommand{\OS}{N}%
\newcommand{\tOS}{\tilde \OS}%
\newcommand{\oOS}{\mathring{\OS}}%
\newcommand{\hvphi}{{\hat\varphi}{}}%
\newcommand{\hrho}{{\hat\rho}{}}%
\newcommand{\etc}{\emph{etc.}}%
\newcommand{\mcP}{{\mycal P}}

\newcommand{\twog}{h}

\newcommand{\beq}{\begin{equation}}

\newcommand{\hq}{\hat q}

\newcommand{\FS}       
                  {F}

\newcommand{\HS} 
       {H_{\mbox{\scriptsize volume}}}

{\ptc{this should be removed in the oberwolfach version}}%

\newcommand{\mcA}{{\mycal A}}%
\newcommand{\eeal}[1]{\label{#1}\end{eqnarray}}
\newcommand{\bed}{\begin{deqarr}}
\newcommand{\eed}{\end{deqarr}}
\newcommand{\bedl}[1]{\begin{deqarr}\label{#1}}
\newcommand{\eedl}[2]{\arrlabel{#1}\label{#2}\end{deqarr}}

\newcommand{\mcU}{{\mycal U}}

\newcommand{\mcN}{{\mycal N}}
\newcommand{\omcN}{\,\,\,\overline{\!\!\!\mcN}}

\newcommand{\bel}[1]{\begin{equation}\label{#1}}
\newcommand{\bea}{\begin{eqnarray}}
\newcommand{\bean}{\begin{eqnarray}\nonumber}
\newcommand{\beal}[1]{\begin{eqnarray}\label{#1}}
\newcommand{\eea}{\end{eqnarray}}

\newcommand{\Eq}[1]{Equation~\eq{#1}}

\newcommand{\qed}{\hfill $\Box$ \medskip}
\newcommand{\proof}{\noindent {\sc Proof:\ }}
\newcommand{\be}{\begin{equation}}
\newcommand{\eeq}{\end{equation}}
\newcommand{\ee}{\end{equation}}
\newcommand{\beqa}{\begin{eqnarray}}
\newcommand{\eeqa}{\end{eqnarray}}
\newcommand{\beqan}{\begin{eqnarray*}}
\newcommand{\eeqan}{\end{eqnarray*}}
\newcommand{\ba}{\begin{array}}
\newcommand{\ea}{\end{array}}

\newcommand{\const}{\mbox{\rm const}} 

\newcommand{\lzeta}{z}

\newtheorem{Theorem} {\sc  Theorem\rm} [section]

\newtheorem{Lemma} [Theorem] {\sc  Lemma\rm}
\newtheorem{Proposition} [Theorem] {\sc  Proposition\rm}

\theorembodyfont{\upshape}
\newtheorem{Remark}[Theorem]{\sc Remark\rm}

\theoremsymbol{\ensuremath{\diamondsuit}}
\newcommand{\fcoco}{\small}
\theorembodyfont{\fcoco} \theoremseparator{} \theoremindent0.5cm

\theoremstyle{nonumberplain} \theorembodyfont{\fcoco}
\theoremseparator{} \theoremindent0.5cm

\DeclareFontFamily{OT1}{rsfs}{}
\DeclareFontShape{OT1}{rsfs}{m}{n}{ <-7> rsfs5 <7-10> rsfs7 <10->
rsfs10}{} \DeclareMathAlphabet{\mycal}{OT1}{rsfs}{m}{n}

{\catcode `\@=11 \global\let\AddToReset=\@addtoreset}
\AddToReset{equation}{section}

\newcounter{mnotecount}[section]

\renewcommand{\themnotecount}{\thesection.\arabic{mnotecount}}

\newcommand{\mnote}[1]
{\protect{\stepcounter{mnotecount}}$^{\mbox{\footnotesize
$
\bullet$\themnotecount}}$ \marginpar{
\raggedright\tiny\em
$\!\!\!\!\!\!\,\bullet$\themnotecount: #1} }

\newcommand{\warn}[1]
{\protect{\stepcounter{mnotecount}}$^{\mbox{\footnotesize
$
\bullet$\themnotecount}}$ \marginpar{
\raggedright\tiny\em $\!\!\!\!\!\!\,\bullet$\themnotecount: {\bf
Warning:} #1} }

\newcommand{\R}{\mathbb R}
\newcommand{\HH}{\mathbb H}

\newcommand{\N}{\mathbb N}

\newcommand{\eq}[1]{(\ref{#1})}

\newcommand{\Mext}{M_\ext}
\newcommand{\ext}{\mathrm{ext}}

\newcommand{\ptc}[1]{\mnote{{\bf ptc:}#1}}

\newcommand{\mcL}{{\mycal L}}

\newcommand{\beqar}{\begin{deqarr}}
\newcommand{\eeqar}{\end{deqarr}}

\newcommand{\beaa}{\begin{eqnarray*}}
\newcommand{\eeaa}{\end{eqnarray*}}

\newcommand{\tr}{\mbox{tr}}
\newcommand{\zg}{\mathring{g}}

\newcommand{\cf}{{\emph{cf.}}}

\newcommand{\hx}{{\hat x}{}}
\newcommand{\hy}{{\hat y}{}}

\newcommand{\eg}{{\emph{e.g.,\/}}}

\title{Mass and angular-momentum inequalities for axi-symmetric initial
data sets
 \\
 I. Positivity of mass}

\author{Piotr T. Chru\'sciel\\%
LMPT, Fédération Denis Poisson,
Tours \\
and \\
Mathematical Institute
,  Oxford}

\begin{document}

\maketitle
\begin{abstract}
We extend the validity of Brill's axisymmetric positive energy
theorem to all asymptotically flat initial data sets with positive
scalar curvature on simply connected manifolds.
\end{abstract}

\tableofcontents

\section{Introduction}
\label{Sintro}


In~\cite{Brill59} Brill proved a positive energy theorem for a
\emph{certain} class of \emph{maximal, axi-symmetric} initial data
sets on $\R^3$. Brill's analysis   has been extended independently
by Moncrief (unpublished){, Dain (unpublished)},  and Gibbons and
Holzegel~\cite{GibbonsHolzegel} to the following class of metrics:
\begin{equation} \label{axmet2}
g = e^{-2U+2\alpha} \left(d\rho^2 + dz^2 \right) + \rho^2 e^{-2U}
\left(d\varphi + \rho B_{\rho} d\rho + A_z dz \right)^2 \, .
\end{equation}
All the functions are assumed to be $\varphi$--independent.

The above form of the metric, together with Brill's formula for the
mass, are the starting points of the recent work of Dain
\cite{Dain:2006}, who proves an upper bound for angular-momentum in
terms of the mass for a class of maximal, vacuum, axi-symmetric
initial data sets with a metric of the form above.

The aim of this series of papers is to extend the validity of
Brill's positivity theorem, as well as that of Dain's inequality, to
all maximal, asymptotically flat, vacuum initial data sets $(M,g,K)$
invariant under a $\Uone$ action on simply connected manifolds.
In fact, our analysis  extends Brill's proof of
positivity of mass to the above class of initial data sets, except
that the condition of being vacuum is weakened to the requirement of
positivity of the matter density.

More precisely, in this paper we prove that any sufficiently
differentiable, asymptotically flat, axially symmetric metric on
$\R^3$ can be written in the form \eq{axmet2}. In general the
functions appearing in \eq{axmet2} will not satisfy the fall-off
conditions imposed in~\cite{GibbonsHolzegel,Dain:2006}, but we
verify that the proof extends to the more general situation. The
result is further extended to include metrics with several
asymptotically flat ends provided the manifold is simply connected.
In the second paper of this series~\cite{ChWY} the constructions of
the current paper will be used to extend the validity of Dain's
angular momentum inequality to the class of metrics considered here.
We will further  allow those non-vacuum models which admit a
\emph{twist potential} $\omega$, see~\cite{ChWY} for details.

It is conceivable that, regardless of simple-connectedness and
isotropy conditions, axi-symmetric metrics on manifolds obtained by
blowing-up a finite number of points in a compact manifold can be
represented as in \eq{axmet2}, with the coordinates $(\rho,z)$
ranging over a subset $\Omega$ of $\R^2$, and with identifications
on $\partial \Omega$, but this remains to be seen; in any case it is
not clear how to adapt the arguments  leading to the mass and
angular momentum inequalities to such situations.%

\section{Axi-symmetric metrics on simply connected asymptotically flat three dimensional manifolds}
\label{Saspet}
 Let us start with a general discussion of Riemannian
manifolds $(M,g)$ with a Killing vector $\eta$ with periodic orbits;
without loss of generality we can assume that the period of
principal orbits is $2\pi$.

Let $\MUone$ denote the collection of the orbits of the group of
isometries generated by $\eta$, and let $\pi:M\to\MUone$ be the
canonical projection. An orbit $p\in \MUone$ will be called
\emph{non-degenerate} if it is \emph{not} a point in $M$. Recall
that near any $p\in \MUone$ which lifts to an orbit of principal
type there exists a canonical metric $q$ defined as follows: let
$X,Y\in T_p(\MUone)$, let $\hat p\in M$ be any point such that $\pi
\hat p =p$, and let $\hat X,\hat Y\in T_{\hat p}M$ be the unique
vectors orthogonal to $\eta$   such that $\pi_* \hat X = X$ and
$\pi_* \hat Y = Y$. Then
\bel{canquotmet}
 q(X,Y) := g(\hat X, \hat Y) \;.
\ee
(The reader will easily check that the right-hand-side of
\eq{canquotmet} is independent of the choice of $\hat p\in
\pi^{-1}(\{p\})$.)

There exists an open dense set of the quotient manifold $\MUone$
which can, at least locally,
be conveniently modeled on smooth submanifolds (perhaps with
boundary), say $\OS$, of $M$, which meet orbits of $\eta$ precisely
once; these are called \emph{cross-sections} of the group action.
(For metrics of the form \eq{axmet2} there actually exists a
\emph{global} cross-section $\OS$, meeting \emph{all} orbits
precisely once.) The manifold structure of $\MUone$ near $p$ is
then, by definition, the one arising from $\OS$. For
$$
 p\in \oOS:=\OS\setminus \{\eta =0\}
$$
and for $X,Y\in T_p\oOS$ set
\bel{OSmet}
 q(X,Y)=g(X,Y)- \frac{g(\eta,X)g(\eta,Y)}{g(\eta,\eta)}
 \;.
\ee
One easily checks that this coincides with our previous definition
of $q$.

The advantage of \eq{OSmet} is that it allows us to read-off
properties of $q$ directly from those of $g$ near $\OS$. On the
other hand,  the abstract definition \eq{canquotmet}  makes clear
the Riemannian character of $q$, and does not require any specific
transverse submanifold. This allows to use different $\OS$'s,
adapted to different problems at hand, to draw conclusions about
$\MUone$; this freedom will be made use of in what follows.

Clearly all the information about $g$ is contained in $q$ and in the
one-form field
$$ \eta^\flat:= g(\eta,\cdot)
\;,
$$
since we can invert \eq{OSmet} using the formula, valid for any
$X,Y\in TM$,
\bel{OSmet2}
  g(X,Y) = q(P_\eta X,P_\eta Y) +
 \frac{g(\eta,X)g(\eta,Y)}{g(\eta,\eta)}
 \;,
\ee
where $P_\eta:TM\to T\oOS$ is the  projection from $TM$ to $T\oOS$
along $\eta$. (Recall that $P_\eta$ is defined as follows: since
$\eta$ is transverse to $T\oOS$, every vector $X\in TM$ can be
uniquely written as $X=\alpha \eta+Y$, where $Y\in T\oOS$, then one
sets $P_\eta X:=Y$.) In order to establish \eq{OSmet2} note, first,
that this is only a rewriting of \eq{OSmet} when both $X$ and $Y$
are tangent to $\oOS$. Next, \eq{OSmet2} is an identity if either
$X$ or $Y$ is proportional to $\eta$, and the result easily follows.

Let $x^A$, $A=1,2$ be any local coordinates on $\oOS$, propagate
them off $\oOS$ by requiring that $\mcL_\eta x^A=0$, and let
$\varphi$ be a coordinate that vanishes on $\oOS$ and satisfies
$\mcL_\eta \varphi = 1$. Then $\eta=\partial_\varphi$, and
$P_\eta(X^A\partial_A+X^\varphi
\partial_\varphi) = X^A\partial_A$, so that \eq{OSmet2} can be rewritten as
\bel{OSmet4}
  g = \underbrace{q_{AB}\,dx^A dx^B}_{q} + {g(\eta,\eta)}(d\varphi +\underbrace{\ttheta _Adx^A}_{=:\ttheta }\;)^2
 \;,
\ee
with
$$
\partial_\varphi q_{AB}= \partial_{\varphi}\ttheta _A=
\partial_\varphi( g(\eta,\eta))=0
 \;.
$$

\subsection{Global considerations}
 \label{sSgc}

So far our considerations were completely general, but  local.
Suppose, however, that $M$ is \emph{simply connected}, with or
without  boundary, and satisfies the usual condition that it is the
union of a compact set and of a finite number of asymptotically flat
ends. Then every asymptotic end can be compactified by adding a
point, with the action of $U(1)$ extending to the compactified
manifold in the obvious way. Similarly every boundary component has
to be a sphere~\cite[Lemma~4.9]{Hempel}, which can be filled in by a
ball, with the action of $U(1)$ extending in the obvious way,
reducing the analysis of the group action to the boundaryless case.
Existence of asymptotically flat regions implies (see,
e.g.,~\cite{ChBeig2}) that the set of fixed points of the action is
non-empty. It is then shown in~\cite{Raymond}    that, after the
addition of a  ball  to every boundary component if necessary, $M$
is homeomorphic to $\R^3$, with the action of $U(1)$ conjugate, by a
homeomorphism, to the usual rotations of $\R^3$. On the other hand,
it is shown in~\cite{Orlik} that the actions are classified, up to
smooth conjugation, by topological invariants. It follows that the
action is in fact smoothly conjugate to the usual rotations of
$\R^3$.  In particular there exists a global cross-section
$\mathring N$ for the action of $U(1)$ away from the set of fixed
points $\mcA$, with $\mathring N$ diffeomorphic to an open
half-plane, with all isotropy groups trivial or equal to $U(1)$, and with $\mcA$ diffeomorphic to $\R$.%
\footnote{I am grateful to Joao Costa and Allen Hatcher for
discussions or comments on the classification of $U(1)$ actions.}

Somewhat more generally, the above analysis applies whenever $M$ can
be compactified by adding a finite number of points or balls. A
nontrivial example  is provided by manifolds with a finite number of
asymptotically flat and asymptotically cylindrical ends, as is the
case for the Cauchy surfaces for the domain of outer communication
of the extreme Kerr solution.

\subsection{Regularity at the axis}
\label{ssRaxis}

 In the coordinates of \eq{axmet2} the   \emph{rotation
axis}
$$
\mcA:=\{g(\eta,\eta)=0\} $$
corresponds to the set $\rho=0$, which for asymptotically flat
metrics is never empty, see, \eg\ the proof of Proposition~2.4 in
\cite{ChBeig2}.

In order to study the properties of $q$ near $\mcA/\Uone\approx
\mcA$, recall that $\mcA$ is a geodesic in $M$. It is convenient to
introduce normal coordinates $(\hx,\hy,\hz):\mcU\to \R^3$  defined
on an open neighborhood $\mcU$ of $\mcA$, where $\hz$ is a
unit-normalized affine parameter on $\mcA$, and $(\hx,\hy)$ are
geodesic coordinates on $\exp((T\mcA)^\perp)$. Without loss of
generality we can assume that $\mcU$ is invariant under the flow of
$\eta$.

As is well known, we have (recalling that orbits of principal type
form an open and dense set of $M$, as well as our normalization of
$2\pi$--periodicity of the principal orbits)
$$
\eta =  \hx \partial_{\hy} - \hy \partial_{\hx}\;.
$$
If we denote by $\phi_t$ the flow of $\eta$, on $\mcU$ the map
$\phi_\pi$ is therefore the  symmetry across the axis $\mcA$:
$$
\phi_\pi (\hx,\hy,\hz) = (-\hx,-\hy,\hz)
 \;.
$$
This formula has several useful consequences. First, it follows that
the manifold with boundary
$$
 \OS:=\{\hx\ge0,\hy=0\}\subset \mcU
$$
is a cross-section for the action of $\Uone$ on $\mcU$. This shows
that near zeros of $\eta$ the quotient space $\MUone$ can be
equipped with the structure of a smooth manifold with boundary. The
analysis of the behavior of $q$ near $\partial \OS\approx \mcA$
requires some work because of the factor $1/g(\eta,\eta)$ appearing
in \eq{OSmet}.

For further use we note that the manifold
\bel{doubleOS}
 \tOS:=  \{ \hy=0\}\subset \mcU
\ee
provides, near $\mcA$, a natural doubling of $\OS$ across its
boundary $\mcA$.

 In order
to understand the smoothness of $q$ on $\OS$ and $\tOS$, we start by
considering the function
$$
 f(\hx,\hz):=g(\eta,\eta)(\hx,0,\hz)
 \;.
$$
Then $f(-\hx,\hz)=f(\hx,\hz)$ because $g(\eta,\eta)\circ \phi_\pi =
g(\eta,\eta)$. It follows that all odd $x$--derivatives of $f$
vanish at $\hx=0$. It is then standard to show, using Borel's
summation lemma (\cf, \eg~\cite[Proposition~C1,
Appendix~C]{ChAnop}), that there exists a smooth function $h(s,\hz)$
such that
$$
f(\hx,\hz)= \hx^2 h(\hx{}^2,\hz)\;.
$$
Letting $\hrho= \sqrt{\hx{}^2+\hy{}^2}$, invariance of $g$ under
$\phi_t$ allows us to conclude that
\bel{etanorm}
g(\eta,\eta)(\hx,\hy,\hz)=g(\eta,\eta)(\hrho,0,\hz)=\hrho^2
h(\hrho^2,\hz)
 \;.
\ee

Define $\hvphi$ via the equations
$$
\hx=\hrho \cos \hvphi\;, \quad \hy = \hrho \sin \hvphi
 \;,
$$
so that
$$
\eta = \partial_\hvphi
 \;.
$$
Considerations similar to those leading to \eq{etanorm}\beijing{
\ptc{I should return to this later}} (see Lemma~5.1 of
\cite{ChAnop}) show that there exist functions $\alpha$, $\beta$,
$\gamma$, $\delta$, $\mu$ and $g_{\hz\hz}$, which are
smooth \emph{with respect to the arguments $\hrho^2$ and $\hz$},%
\footnote{By this we mean that $\alpha(s,\hz)$ is a smooth function
of its arguments, and enters \eq{gaxism37} in the form
$\alpha(\hrho^2,\hz)$, \etc}
with
$$\mu(0,\hz)=1\;, \quad g_{\hz\hz}(0,\hz) =1
 \;,
$$
such that
\bean
 g &=& g_{\hz\hz}d\hz^2+2 \alpha \hrho d\hz d\hrho
+2\beta \hrho^2 d\hz d\hvphi + \gamma\hrho^2 d\hrho^2 +2 \delta
\hrho^3 d\hrho d\hvphi+ \mu(d\hrho^2+\hrho^2 d\hvphi^2)
 \\
 & = &
  \underbrace{\Big(g_{\hz\hz}- \frac{\beta^2\hrho^2}\mu\Big)d\hz^2+2 \Big(\alpha - \frac{\delta\beta\hrho^2}\mu \Big)\hrho d\hz d\hrho
+\Big(\mu +\gamma\hrho^2- \frac{\delta^2\hrho^2}\mu\Big)
d\hrho^2}_{q}
 \nonumber
 \\
 &  &
+ \mu\hrho^2\Big(d\hvphi+\underbrace{\frac\delta\mu \hrho
d\hrho+\frac\beta\mu d\hz}_{\ttheta }\;\Big)^2 \;.
 \eeal{gaxism37}

We say that  $\hat N$ is a \emph{doubling} of a manifold $N$ across
a boundary $\dot N$ if $\hat N$ consists of two copies of $N$ with
points on $\dot N$ identified in the obvious way. From what has been
said, by inspection of \eq{gaxism37} it follows that:

\begin{Proposition}
\label{Paxsymgr} The quotient space $\MUone$ has a natural structure
of manifold with boundary near $\mcA$. The metric $q$ and the
one-form $\ttheta $ are smooth up-to-boundary, and extend smoothly
across $ \mcA$ by continuity to themselves when $\MUone$ is doubled
at $\mcA$.
\end{Proposition}

For further use we note the formula
\bel{gphirel}
 g(\eta,\eta) = \hrho^2 + O(\hrho^4)\;,
\ee
for small $\hrho$, which follows from \eq{gaxism37}, where $\hrho$
is either the geodesic distance from $\mcA$, or the geodesic
distance from $\mcA$ on $\exp((T\mcA)^\perp)$ (the latter being, for
small $\hrho$, the restriction to $\exp((T\mcA)^\perp)$ of the
former).

\subsection{Asymptotic flatness}
\label{ssAf}

We will consider Riemannian manifolds $(M,g)$ that are
asymptotically flat, in the usual sense that there exists a region
$\Mext\subset M$ diffeomorphic to $\R^3\setminus B(R)$, where $B(R)$
is a coordinate ball of radius $R$, such that in local coordinates
on $\Mext$ obtained from $\R^3\setminus B(R)$ the metric satisfies
the fall-off conditions, for some $k\ge 1$,
\beal{falloff1}
 & g_{ij}-\delta_{ij}=o_k(r^{-1/2})\;,
  \\
  & \label{falloff1a}
\partial_k g_{ij}\in L^2(\Mext)\;,
 &
 \\
 &R^i{}_{jk\ell}=o(r^{-5/2})\;,
 &
 \eeal{falloff2}
where we write $f=o_k(r^{\alpha})$ if $f$ satisfies
$$
  \partial_{k_1}\ldots\partial_{k_\ell}
f=o (r^{\alpha-\ell})\;, \quad 0\le \ell \le k
 \;.
$$
It is well known that \eq{falloff1}-\eq{falloff1a} together with
$R(g)\ge 0$ or $R(g)\in L^1$, where $R(g)$ is the Ricci scalar of
$g$, guarantees a well-defined ADM mass (perhaps infinite). On the
other hand, the condition \eq{falloff2} (which follows in any case
from \eq{falloff1} for $k\ge 2$) is useful when analyzing the
asymptotic behavior of Killing vector fields.

We will use \eq{falloff1}-\eq{falloff2} to construct the coordinate
system of \eq{OSmet2}, and also to derive the asymptotic behavior of
the fields appearing in \eq{OSmet2}.   We start by noting that the
arguments of~\cite[Appendix~C]{ChBeig1} with $N\equiv 0$ there show
that there exists a rotation matrix $\omega$ such that in local
coordinates on $\Mext$ we have
\bel{roughasym}
 \eta^i= \omega^i{}_j x^j + o_k(r^{1/2})
 \;,
\ee
where $\omega^i{}_j $ is anti-symmetric. It will be clear from the
proof below (see \eq{extraterms}) that this equation provides the
information needed in the region
\bel{goodregion}
 x^2+y^2\ge z^2\;, \quad x^2+y^2+z^2\ge R^2
\;. \ee
However, near the axis a more precise result is required, and we
continue by constructing new asymptotically flat coordinates which
are better adapted to the problem at hand. The difficulties arise
from the need to obtain decay estimates on $q-\delta$, where $\delta
$ is the Euclidean metric on $\R^2$, and on $\ttheta $, which are
\emph{uniform in $r$ up to the axis} $\mcA$.

Let $(\hat x^i)\equiv (\hat x,\hat y,\hat z)$ be coordinates on
$\R^3\setminus B(R)$, obtained by a rigid rotation of $x^i$, such
that $\omega^i{}_j \hat x^j =\hat y\partial_{\hx} - \hx
\partial_{\hy}$. Set
\bea
 &
 \displaystyle
x:= \frac{\hx-\hx\circ\phi_\pi}2 
 \;, \quad
 y:= \frac{\hy-\hy\circ\phi_\pi}2
 \;, \quad
 z:= \frac 1 {2\pi} \int_0^{2\pi} \hat z\circ \phi_s \, ds\;.
 &
\eeal{xdef2}
 Using the techniques in~\cite{ChBeig1,ChBeig2}
one finds
$$
 \phi_s(\hx^i)=(\cos(s)\hx-\sin(s)\hy+\lzeta ^\hx(s,\hx^i),
 \sin(s)\hx+\cos(s)\hy+\lzeta ^\hy(s,\hx^i),\hz+\lzeta ^\hz(s,\hx^i))
 \;,
$$
with $\lzeta ^i$ satisfying
$$
 \lzeta ^i = o_{k+1}(r^{1/2})
 \;.
$$
We then have
\beaa
 \frac{\partial z}{\partial \hat z}
 & = & 1 +
 \frac 1 {2\pi} \int_0^{2\pi} \frac{\partial \lzeta ^{\hat z}(\phi_s(\hx^i))}{\partial \hz} \,ds
 =
  1 + o_{k}(r^{-1/2}) \;, \eeaa
Further,
\beaa
 \frac{\partial z}{\partial \hat x}&=&
 \frac 1 {2\pi} \int_0^{2\pi} \frac{\partial \lzeta ^{\hat z}(\phi_s(\hx^i))}{\partial \hx} \,ds
=
   o_{k}(r^{-1/2}) \;, \eeaa
similarly
\beaa
 \frac{\partial z}{\partial \hat y}&=&
   o_{k}(r^{-1/2}) \;.
\eeaa
The estimates for the derivatives of $x$ and $y$ are
straightforward, and we conclude that
$$
\frac{\partial x^i}{\partial \hat x^i} = \delta^i_j +o_{k}(r^{-1/2})
\;,
$$
where, by an abuse of notation, we write again $x^i$ for the
functions $(x,y,z)$. Standard considerations based on the implicit
function theorem show that, increasing $R$ if necessary, the $x^i$'s
form a coordinate system on $\R^3\setminus B(R)$ in which
\eq{falloff1}-\eq{falloff2} hold. Subsequently, \eq{roughasym} holds
again.

From  \eq{xdef2} one clearly has
$$
\forall \ s\in \R \qquad
 z\circ \phi_s = z\;,
$$
%
which shows that the planes
$$
 \mcP_\tau:=\{z=\tau\}\;, \quad \tau \in \R \;, \ |\tau|\ge R
 \;,
$$
are invariant under the flow of $\eta$; equivalently,
$$
 \eta^z=0
 \;.
$$
Moreover,
\bel{xyinvInv}
 x\circ \phi_\pi = -x\;, \quad
 y\circ \phi_\pi = -y\;,
\ee
so that all  points with coordinates $x=y=0$ are fixed points of
$\phi_\pi$, and that \emph{these are the only such points in
$\Mext$}.   \Eq{xyinvInv} further implies that $\phi_\pi$ maps the
surfaces $\{x=0\}$ and $\{y=0\}$ into themselves. Since $\phi_\pi$
is an isometry, we obtain
\bean
 &
 g_{ab}(0,y,z)=g_{ab}(0,-y,z)\;,\quad
 g_{zz}(0,y,z)=g_{zz}(0,-y,z)\;,
 &
 \\
 &
 g_{za}(0,y,z)=-g_{za}(0,-y,z)\;;
 &
\eeal{xyinvInv2}
similarly
\bean
 &
 g_{ab}(x,0,z)=g_{ab}(-x,0,z)\;,\quad
 g_{zz}(x,0,z)=g_{zz}(-x,0,z)\;,
 &
 \\
 &
 g_{za}(x,0,z)=-g_{za}(-x,0,z) \;.
 &
 \eeal{xyinvInv3}
\Eq{xyinvInv2} leads to
\bel{xyinvINV4}
 \frac{\partial^{2\ell+1}g_{ab}}{\partial
y^{2\ell+1}}(0,0,z)=0
 \;,
 \quad
  \frac{\partial^{2\ell+1}g_{zz}}{\partial
y^{2\ell+1}}(0,0,z)=0
 \;,
 \quad
  \frac{\partial^{2\ell }g_{az}}{\partial
y^{2\ell }}(0,0,z)=0
 \ee
for $\ell\in \N$ (or at least as far as the differentiability of the
metric allows). The analogous implication of \eq{xyinvInv3} allows
us to conclude that
%
\bel{axdercontrol}
 \frac{\partial g_{ab}}{\partial x^{c }}(0,0,z)=0
 \;, \quad
 \frac{\partial g_{zz}}{\partial x^{a }}(0,0,z)=0
 \;, \quad
  g_{az} (0,0,z)=0
 \;.
\ee
Incidentally, the last two equations in \eq{axdercontrol} show that
$\{x=y=0\}$ is a geodesic; this follows in any case from the
well-known fact that the set of fixed points of an isometry is
totally geodesic.

Consider a point $p$ lying on the axis of rotation $\mcA$, then
$\phi_t(p)=p$ for all $t$, in particular $\phi_\pi(p)=p$. From what
has been said we obtain that
\bel{goodinclusion}
 \mcA\cap \Mext \subset\{x=y=0\}
 \;.
\ee
  Recall, again, that every connected component of the axis of rotation $\mcA$
is \emph{an inextendible geodesic} in $(M,g)$. Since the set at the
right-hand-side of \eq{goodinclusion} is a geodesic segment, we
conclude that equality holds in \eq{goodinclusion}. Hence
\bel{vanishingeta}
 \eta^i(0,0,z)=0
\ee
and, for $|z|\ge R$, the origin is the only point within the plane
$\mcP_z$ at which $\eta$ vanishes.

We are ready now to pass to the problem at hand, namely an
asymptotic analysis of the fields $g(\eta,\eta)$, $q$ and $\ttheta $
as in \eq{OSmet4}; we start with $q $. For $\rho$ sufficiently large
the hypersurface $\{y=0\}$ is transverse to $\eta$ (for small $\rho$
we will return to this issue shortly) and therefore the coordinates
$$
 (x^A):=(x,z)
 $$
on this hypersurface, with $x\ge 0$, can be used as local
coordinates on $\MUone$. The contribution of $g_{AB}$ to $q_{AB}$ is
of the form $g_{AB}=\delta_{AB}+o_{k}(r^{-1/2})$, which is
manifestly asymptotically flat in the usual sense. Next, from
\eq{falloff1} and \eq{roughasym} we obtain
\bel{getaeta} g(\eta,\eta)= \rho^2 + o_{k}(r^{3/2})
 \;;
\ee
here, as elsewhere, $\rho^2=x^2+y^2$. Further
\bean
 \frac{g_{Ai}\eta^i g_{Bj}\eta^j}{g(\eta,\eta)}
 dx^A dx^B
  &= &
  \Big(\delta_{Ai}
 +o_{k}(r^{-1/2})\Big)\Big(\omega^i{}_a x^a +
 o_{k}(r^{1/2})\Big)
 \times
 \\
 \nonumber
 &&
 \frac{
 \Big(\delta_{Bj}
 +o_{k}(r^{-1/2})\Big)\Big(\omega^j{}_b x^b +
 o_{k}(r^{1/2})\Big)}
 {
 \rho^2 + o_{k}(r^{3/2})
 }
 dx^A dx^B
 \\
 &=& \frac{  o_{k}(r^{1/2})
 dx^A dx^B}
 {
 \rho^2 + o_{k}(r^{3/2})
 }
 \;,
\eeal{extraterms}
because $ \omega^i{}_a x^a \omega^j{}_b x^b dx^i dx^j =
(xdy-ydx)^2$, which vanishes when pulled-back to $\{y=0\}$. In the
region \eq{goodregion} we thus obtain
\bel{qasym}
 q_{AB} = \delta_{AB}+o_{k }(r^{-1/2})
 \;,
 \ee
which is the desired estimate. However,  near the zeros of $\eta$
this calculation is not precise enough to obtain uniform estimates
on $q$ and its derivatives.

In fact, it will be seen in the remainder of the proof that we need
uniform estimates for derivatives up to second order. Since
$g(\eta,\eta)$ vanishes quadratically at the origin we  need uniform
control of the numerator of \eq{extraterms} up to terms $O(\rho^4)$,
in a form which allows the division to be performed without losing
uniformity.

So in the region $\{\rho \le |z|\}\cap \Mext$, in which $|z|$ is
comparable with $r$, we proceed as follows: Let
$$
 \lambda^a{}_b\equiv\lambda^a{}_b(z):= \frac{\partial\eta^a}{\partial x^b}
 (0,0,z)\;, \quad \lambda_{ab}:= g_{ac}(0,0,z)\lambda^c{}_b
 \;;
$$
note that
$\lambda^a{}_b=\omega^a{}_b+o_{k-1}(|z|^{-1/2})=\omega^a{}_b+o_{k-1}(r^{-1/2})$,
similarly for $\lambda_{ab}$. The Killing equations imply that
$\lambda_{ab}$ is anti-symmetric, hence
$$
\lambda_{xx}=\lambda_{yy}=0\;, \qquad
 \lambda_{xy}=-\lambda_{yx}=1+ o_{k-1}(|z|^{-1/2})=1+o_{k-1}(r^{-1/2})
 \;.
$$
From \eq{vanishingeta} we further obtain
$$
 \partial_i \eta^z=0 \quad \Longrightarrow \quad \nabla_i \eta^z|_\mcA=0
 \quad \Longrightarrow \quad   \nabla_i \eta_z|_\mcA= \nabla_z
 \eta_i|_\mcA=\nabla_z
 \eta^i|_\mcA=0
 \;.
$$

Recall the well known consequence of the Killing equations,
$$
 \nabla_i \nabla_j \eta_k = R^\ell{}_{ijk}\eta_\ell
 \;,
$$
which implies, at $\mcA$,
\beal{secderv1}
  &
  0 = \nabla_a \nabla_b \eta_c=\partial_a \partial_b \eta_c\;,
  &
  \\
  &  0 =\nabla_a \nabla_b \eta_z=\partial_a \nabla_b
\eta_z-\Gamma^c{}_{az} \lambda_{bc}=\partial_a \partial_b
\eta_z-2\Gamma^c{}_{az} \lambda_{bc}
 \;.
 &
\eeal{secderv2}
%
%
From \eq{roughasym} we obtain, by integration of third derivatives
of $\eta_a$ along rays from the origin $x=y=0$ within the planes
$z=\const$,
$$
 \frac{\partial^2\eta_a}{\partial x^b\partial x^c} =    o_{k-3}(|z|^{-5/2})x^c
 =  o_{k-3}(r^{-5/2})x^c
 \;,
$$
and then successive such integrations give
$$
 \frac{\partial\eta_a}{\partial x^b} =  \lambda_{ab} +
 o_{k-3}(|z|^{-5/2})x^cx^d
 = \lambda_{ab}  + o_{k-3}(r^{-5/2})x^cx^d
 \;,
$$
\bel{etaest1}
 \eta_a =  \lambda_{ab}x^b + o_{k-3}(r^{-5/2})x^cx^dx^e
 \;.
\ee At $y=0$ we conclude that
$$
\eta_x=  o_{k-3}(r^{-5/2})x^cx^dx^e
 \;.
 $$
 Similarly we have    $
 \nabla_i \nabla_j \eta^k = R^\ell{}_{ij{}}{}^{k}\eta_\ell
 $, hence  $ \nabla_a \nabla_b \eta^c=\partial_a \partial_b
 \eta^c=0$  at $\mcA$,
 and we conclude that
\bel{etaest1b}
 \eta^a =  \lambda^a{}_{b}x^b + o_{k-3}(r^{-5/2})x^cx^dx^e
 \;.
\ee
This allows us to prove transversality of $\eta$ to the plane
$\{y=0\}$. Indeed, from \eq{etaest1b} at $y=0$ we have
$$
 \eta^y = (1+o (r^{-1/2}))x +o (r^{-5/2})x^3=(1+o (r^{-1/2}))x
$$
%
which has no zeros for $x\ne 0$ and $r\ge R$ if $R$ is large enough.
Recall that we have been assuming that $|x| \le |z|$ in the current
calculation; however, we already know that $\eta$ is transverse for
$|z|\ge |x|$,  and transversality follows. Increasing the value of
the radius $R$ defining $\Mext$ if necessary, we conclude that
$\{y=0, x\ge 0\}\cap \Mext$ provides a global cross-section for the
action of $\Uone$ in $\Mext$.

Using \eq{secderv2}, a similar analysis of $\eta_z$ gives
$$
\eta_z=-\underbrace{\Gamma^c{}_{az}|_\mcA}_{o_{k-1}(r^{-3/2})}
\lambda_{bc} x^a x^b +o_{k-3}(r^{-5/2})x^cx^dx^e
 \;.
$$

We are now ready to return to \eq{getaeta},
\bea
 g(\eta,\eta)= \eta_i \eta^i= \eta_a \eta^a =
 \hat \rho^2 +o_{k-3}(r^{-5/2})x^{a }x^{b}x^{c }x^d
 \;,
\eeal{getaeta2}
where, at $y=0$,
$$
 \hat \rho^2:= \zg_{ab}\lambda^a{}_c x^c\lambda^b{}_d x^d = (1+
o_{k-1}(r^{-1/2}))x^2
 \;;
$$
it follows that the last equality also holds for $g(\eta,\eta)$ with
$k-1$ replaced by $k-3$. Instead of \eq{extraterms} we write
\bean
 \frac{g_{Ai}\eta^i g_{Bj}\eta^j}{g(\eta,\eta)}
 dx^A dx^B
  &= &
 \frac{\eta_A \eta_B
 dx^A dx^B}
 {
 (1+
o_{k-3}(r^{-1/2}))x^2
 }
 \\
 &=& \frac{ \eta_x^2\,
 dx^2+ 2\eta_x \eta_z
 dx \,dz + \eta_z^2\, dz^2}
 {
 (1+
o_{k-3}(r^{-1/2}))x^2
 }
  \nonumber
   \\
 &=&  \frac{o_{k-3}(r^{-3})x^2 dx^A dx^B}
 {
 (1+
o_{k-3}(r^{-1/2}))
 }
  \nonumber
   \\
 &=&o_{k-3}(r^{-1})  dx^A dx^B
  \;.\phantom{xxxxx}
\eeal{extraterms2}
We conclude that \eq{qasym} holds throughout $\{y=0\}\cap\Mext$ with
$k$ replaced by $k-3$.

To analyse the fall-off of $B_\rho$ and $A_z$, note first that the
discussion in the paragraph before \eq{OSmet4} shows that it
suffices to do this at one single surface transverse to the flow of
the Killing vector field $\eta$; unsurprisingly, we choose
$$
 N:=\{y=0\,,\ x>0\,,\ x^2+z^2\ge R^2 \}
 \;,
$$
with $R$ sufficiently large to guarantee transversality. Next, from
\eq{axmet2} we find
$$
\eta_i dx^i = g(\eta, \cdot) = g(\partial_\varphi, \cdot) =
g(\eta,\eta)(d\varphi + \rho B_\rho d\rho+A_z dz)
 \;,
$$
which will allow us to relate $B_\rho$ and $A_z$ to $\eta_i$ if we
determine, say $\partial_i \varphi$ and $\partial_ i \rho$ on $N$.
For the sake of clarity of intermediate calculations it is
convenient to denote by $\bar z$ the coordinate $z$ appearing in
\eq{axmet2}, we thus seek a coordinate transformation
$$
 (x,y,z) \to (\rho,\varphi,\bar z)\;, \ \mbox{with $\bar z = z $
 everywhere and $\rho=x$ on $N$},
$$
which brings the metric to the form \eq{axmet2}, with $z$ there
replaced by $\bar z$. We wish to show that, on $N$,
\bel{Jacob}
 J:=\left(
\begin{array}{ccc}
  \frac{\partial x}{\partial \rho} & \frac{\partial x}{\partial \varphi} & \frac{\partial x}{\partial \bar z} \\
  \frac{\partial y}{\partial \rho} & \frac{\partial y}{\partial \varphi} & \frac{\partial y}{\partial \bar z} \\
  \frac{\partial z}{\partial \rho} & \frac{\partial z}{\partial \varphi} & \frac{\partial z}{\partial \bar z} \\
\end{array}
\right)
 =
  \left(
\begin{array}{ccc}
  1 & \eta^x & 0 \\
  0 & \eta^y & 0 \\
  0 & 0 & 1
\end{array}
 \right)
 \;.
\ee
The second column is immediate from
$$
\eta^x\partial_x + \eta^y \partial_y + \eta^z \partial_z=\eta=
\partial_\varphi= \frac{\partial x}{\partial \varphi}\partial_ x +
  \frac{\partial y}{\partial \varphi}\partial_y
  +
  \frac{\partial z}{\partial \varphi}\partial_ z
  \;.
$$
Similarly the third row follows immediately from $dz=d\bar z$. It
seems that the remaining entries require considering $J^{-1}$. Now,
$\varphi$ is a coordinate that vanishes on $N$, so that
$\partial_x\varphi=\partial_z\varphi=0$ there. From $\eta^i
\partial_i \varphi=1$ we thus obtain
$\partial_y\varphi=1/\eta^y$. Next, $\rho=x$ on $N$, giving
$\partial_x\rho=1$ and $\partial_z\rho=0$ there. The equation
$\eta^i \partial_i\rho =0$ gives then $\eta^x+\eta^y \partial_y
\rho=0$, so that $\partial_y\rho=-\eta^x/\eta^y$. The derivatives of
$\bar z$ are straightforward, leading to
$$
 J^{-1}=\left(
\begin{array}{ccc}
  \frac{\partial \rho}{\partial x} & \frac{\partial \rho}{\partial y} & \frac{\partial \rho}{\partial   z} \\
  \frac{\partial \varphi}{\partial x} & \frac{\partial \varphi}{\partial y} & \frac{\partial \varphi}{\partial  z} \\
  \frac{\partial \bar z}{\partial x} & \frac{\partial \bar z}{\partial y} & \frac{\partial \bar z}{\partial z} \\
\end{array}
\right)
 =
  \left(
\begin{array}{ccc}
  1 & -\eta^x/\eta^y & 0 \\
  0 & 1/\eta^y & 0 \\
  0 & 0 & 1
\end{array}
 \right)
 \;.
$$
Inverting $J^{-1}$ leads to \eq{Jacob}.

From now on we drop the bar on $\bar z$. From \eq{Jacob} one
immediately has on $N$
\bean
 A_z &=&  \frac {\eta_z}{g(\eta,\eta)}=\left\{
                                       \begin{array}{ll}
                                       o_{k-1}(r^{-3/2})+o_{k-3}(r^{-5/2})x, & \hbox{$|x| \le |z|$}, \\
                                         o_{k}(r^{-3/2}), &
\hbox{otherwise},
                                       \end{array}
                                     \right.
 \\
 & = &
 o_{k-3}(r^{-3/2})
 \;.
\eeal{Afof}
Similarly, again on $N$,
\bean
  B_\rho &=& \frac{\eta_i}{\rho g(\eta,\eta)}\frac{\partial x^i}
{\partial \rho}
 = \frac{\eta_x}{x g(\eta,\eta)}
 =\left\{
                                       \begin{array}{ll}
                                         o_{k-3}(r^{-5/2}), & \hbox{$|x|  \le |z|$}, \\
                                         o_{k}(r^{-5/2}), &
\hbox{otherwise},
                                       \end{array}
                                     \right.
 \\
 & = &
 o_{k-3}(r^{-5/2})\;.
 \eeal{Brho}
Finally, we note that
\bean
  e^{-2U}&:=& \frac{g(\eta,\eta)}{\rho^2} =\left\{
                                       \begin{array}{ll}
                                        1+
o_{k-1}(r^{-1/2})  + o_{k-3}(r^{-5/2})x^2, & \hbox{$|x|  \le |z|$}, \\
                                        1+ o_{k}(r^{-1/2}), &
\hbox{otherwise},
                                       \end{array}
                                     \right.
 \\
 & = &
 1+o_{k-3}(r^{-1/2})\;.
 \eeal{Brho2}

In summary:

\begin{Proposition}
\label{PAxsymaf} Under \eq{falloff1} with $k\ge 3$
the metric $q$ is asymptotically flat. In fact, there exist
coordinates $(x,y,z)$ satisfying \eq{falloff1} and a constant
$R\ge 0$ such that the plane $\{y=0\}\cap \{r\ge R\}$ is
transverse to $\eta$ except at $x=z=0$ where $\eta$ vanishes and,
setting $x^A=(x,z)$ we have
\bel{twodimAFq}
 q_{AB}-\delta_{AB} = o_{k-3}(r^{-1/2})
 \;.
\ee
Furthermore \eq{Afof}-\eq{Brho2} hold.
\end{Proposition}

\subsection{Isothermal coordinates}

We will use the same symbol $q$ for metric on the manifold obtained
by doubling $\MUone$ across the axis.

We start by noting the following:

\begin{Proposition}
\label{Pisocor} Let $q$ be an asymptotically flat metric on $\R^2$
in the sense of \eq{twodimAFq} with $k\ge 5$. Then $q$ has a global
representation
\bel{vwcoords} q=e^{2u}(dv^2+dw^2)\;, \quad \mbox{with} \
u\longrightarrow_{\sqrt{v^2+w^2} \to \infty}0\;. \ee
In fact, $u=o_{k-4}(r^{-1/2})$.
\end{Proposition}

\begin{Remark}
The classical justification of the existence of global isothermal
coordinates proceeds by constructing the coordinate $v$ of
\eq{vwcoords} as a solution of the equation $\Delta_q v=0$. A more
careful version of the arguments in the spirit
of~\cite[Lemma~2.3]{MzHS2} shows that $v$ has no critical points.
However, the approach here appears to be  simpler.
\end{Remark}

\proof Let $\tilde q_{AB}=e^{-2u} q_{AB}$, then $\tilde q$ is flat
if and only if $u$ satisfies the equation
\bel{utwoeq}
 \Delta_q u = -\frac {R(q)}2
 \;,
\ee
where $R(q)$ is the scalar curvature of $q$. For asymptotically
flat metrics $q$, with asymptotically Euclidean coordinates
$(x,z)$, this equation always has a solution such that
\bel{mudefR}
 u + \mu \ln (\sqrt{x^2+z^2})\longrightarrow_{\sqrt{x^2+z^2}
\to \infty}0\;, \quad \mbox{\rm where} \ \; \mu
 = \frac 1{4\pi} \int_{\R^2} R(q) \,d\mu_q  \;,
\ee
where $d\mu_q$ is the volume form of $q$.
More precisely, we have the following:

\begin{Lemma}
\label{LExLapl} Consider a metric $q$ on $\R^2$ satisfying
$$
q_{AB}-\delta_{AB}=o_\ell(r^{-1/2})
$$
for some $\ell \ge 2$, with $(x^A)=(x,z)$. For any continuous
function $R=o_{\ell-2}(r^{-5/2})$ there exists $\hat
u=o_{\ell-1}(r^{-1/2})$ and a solution of \eq{utwoeq} such that
$$u
 =\hat u-\mu \ln
 (\sqrt{x^2+z^2})
 \;,
$$
with $\mu$ as in \eq{mudefR}.
\end{Lemma}

\proof
We start by showing that \eq{utwoeq} can be solved for $|x|$ large.
Indeed, consider the sequence $v_i$ of solutions of \eq{utwoeq} on
the annulus
$$ \Gamma(\rho,\rho+i):= D(0,\rho+i)\setminus D(0,\rho)
 \;,
$$
with zero boundary values. Here $\rho$ is a constant chosen large
enough so that the functions $\pm C|x|^{-1/2}$, with
$C=8\|R|x|^{5/2}\|_{L^\infty}$, are sub- and super-solutions of
\eq{utwoeq}. The usual elliptic estimates show that a subsequence
can be chosen which converges, uniformly on compact sets, to a
solution $v=O_{\ell-1}(r^{-1/2})$ of \eq{utwoeq} on $\R^2\setminus
D(0,\rho)$. In the notation of~\cite{ChAFT} we have in fact $v \in
C^{\ell-1,\lambda}_{-1/2,0}$ for any $\lambda \in (0,1)$.
Furthermore, using the techniques in~\cite{ChAFT} one checks that
$v=o_{\ell-1}(r^{-1/2})$.

We extend $v$ in any way to a $C^{\ell-1,\lambda}$ function on
$\R^2$, still denoted by $v$. Let $\hq:=e^{-2v} q$, then $\hq$ is
flat for $|x|\ge \rho$. Let $\hat e^A$ be any $\hq$--parallel
orthonormal co-frame on $\R^2\setminus D(0,\rho)$, performing a
rigid rotation of the coordinates if necessary we will have $\hat
e^A= dx^A+\sum_Bo_{\ell-1}(|x|^{-1/2})dx^B$ for $|x|$ large. Let
$\hx^A$ be any solutions of the set of equations $d\hat x^A=\hat
e^A$. By the implicit function theorem the functions $\hat x^A$
cover $\R^2\setminus D(0,\hat \rho)$, for some $\hat \rho$, and form
a coordinate system there, in which $\hq_{AB}=\delta_{AB}$.

Since \eq{utwoeq} is conformally covariant, we have reduced the
problem to one where $R$ has compact support, and $q$ is a
$C^{\ell-1,\lambda}$ metric which is flat outside of a compact set.
This will be assumed in what follows.

Let us use the stereographic projection, say $\psi$, to map $\R^2$
to a sphere, then \eq{utwoeq} becomes an equation for $\hat u:= (u
-v)\circ \psi^{-1}$ on $S^2\setminus\{i^0\}$, where $i^0$ is the
north pole of $S^2$, of the form
\bel{utwoeq2}
 \Delta_\twog \hat u = |x|^4 f
 \;,
\ee
where $\twog_{AB}:=|x|^{-4}q_{AB}$ is a $C^{\ell-1,\lambda}$ metric
on $S^2$, similarly $f$ is a  $C^{\ell-2}$ function on $S^2$
supported away from the north pole. In fact, in a coordinate system
\bel{inversionform}
 y^A=x^A/|x|^2
\ee
near $i^0=\{y^A=0\}$, where the $x^A$'s are the explicitly flat
coordinates  on $\R^2\setminus D(0,R)$ for the metric $q$, we have
$$
 \twog_{AB}=\delta_{AB}
\;.
$$
%

Let $H_k(S^2)$ be the usual $L^2$-type Sobolev space of functions on
$S^2$ and set
\bel{Hkdef}
 \HH_k = \Big\{\chi \in H_k(S^2)\ | \ \int_{S^2} \chi \,d\mu_\twog = 0 \Big\}
 \;,
\ee
where $d\mu_\twog$ is the measure associated with the metric
$\twog$. We have

\begin{Proposition}
\label{Piso} Let $h$ be a twice-differentiable metric on $S^2$, then
$\Delta_\twog:\HH_2\to \HH_0$ is an isomorphism.
\end{Proposition}

\proof Injectivity is straightforward. To show surjectivity, let
$X\subset L^2$ be the image of $\HH_2$ by $\Delta_\twog$, by
elliptic estimates $X$ is a closed subspace of $L^2(S^2)$. Let
$\varphi\in L^2$ be orthogonal to $X$, then
$$
\forall \chi\in \HH_2 \quad \int \varphi \Delta_\twog \chi  \,
d\mu_\twog= 0
 \;.
$$
Thus $\varphi$ is a weak solution of $\Delta_\twog \varphi=0$, by
elliptic estimates  $\varphi\in \HH_2$. But setting $\chi=\varphi$
and integrating by parts one obtains $d\varphi=0$, hence $\varphi$
is constant, which shows that $X=\HH_0$.
 \qed

Returning to the proof of Lemma~\ref{LExLapl}, we have seen that
\eq{utwoeq} can be reduced to solving the problem
\bel{eq33}
 \Delta_{\bar\twog} \bar u= \bar f
 \;,
\ee
where $\bar \twog$ is flat outside of a compact set. Let
$$
 \mu := -\frac 1 {2\pi} \int_{\R^2}\bar f\, d\mu_{\bar \twog}
 \;,
$$
then
\beaa
 \int_{\R^2} \Delta_{\bar\twog}\Big( \mu \ln \sqrt{1+x^2+z^2}\Big)\,d\mu_{\bar \twog} &=& \lim_{\rho\to\infty}
 \mu\oint_{C(0,\rho)} D\Big(\ln \sqrt{1+x^2+z^2}\Big) \cdot n
\\
 &  = & 2\pi \mu= -\int_{\R^2}\bar f\, d\mu_{\bar \twog}
\eeaa
Thus \eq{eq33} is equivalent to the following equation for the
function $\tilde u:= \bar u + \mu \ln \sqrt{1+x^2+z^2}$:
$$
\Delta_{\bar\twog} \tilde u = \bar f +\Delta_{\bar\twog}\Big( \mu
\ln \sqrt{1+x^2+z^2}\Big)
 \;,
$$
and the right-hand-side has vanishing average. Transforming to a
problem on $S^2$ as in \eq{utwoeq2}, we can solve the resulting
equation by Proposition~\ref{Piso}. Transforming back to $\R^2$, and
shifting $u$
by a constant if necessary, the result follows.%
%
\qed

Returning to the proof of Proposition~\ref{Pisocor}, we claim that
$\mu=0$; that is,
\bel{2curvcond}
 \int_{\R^2} R(q) \,d\mu_q = 0 \;.
\ee
This is the simplest version of the Gauss-Bonnet theorem, we give
the proof for completeness: consider any metric on $ \R^2$
satisfying
$$
 q_{AB}-\delta_{AB} = o_{1}(1)
 \;, \qquad R(q)\in L^1 \;.
$$
Let $ \ttheta ^a$, $a=1,2$, be an orthonormal co-frame for $q$
obtained by a Gram-Schmidt procedure starting from $(dx^1, dx^2)$,
with connection coefficients $\omega^a{}_b$. Then
$\omega^a{}_b=o(r^{-1})$. It is well known that, in dimension two,
\bel{Rid}
 R(q)\, d\mu_q = 2d\omega^1{}_2
 \;.
\ee
\Eq{2curvcond} immediately follows by integration on $B(R)$, using
Stokes' theorem, and passing to the limit $R\to \infty$.

To finish the proof, note that the metric $\tilde q$ is a complete
flat metric on $\R^2$, and the Hadamard-Cartan theorem shows the
existence of global manifestly flat coordinates, say $(v,w)$ so that
$q$  can be written as in \eq{vwcoords}.
\qed

Returning to the problem at hand, recall that the metric $q$ on
$\R^2$ has been obtained by doubling $\MUone$ across $\mcA$. Let
us denote by $\phi$ the corresponding isometry; note that in
$\Mext/\Uone$, in the coordinates $(x,z)$ constructed in
Section~\ref{ssAf}, the isometry $\phi$ is the symmetry around the
$z$-axis: $\phi(x,z)= (-x,z)$. Similarly, in  geodesic coordinates
centred on $\mcA$, $\phi(x,z)= (-x,z)$.

As $\phi$ is an isometry of $q$, preserving the boundary conditions
satisfied by $u$, uniqueness of solutions of \eq{utwoeq} implies
that $u\circ \phi=u$. Smoothness on the doubled manifold shows that
on $\mcA$ the gradient $\nabla u$ has only components tangential to
$\mcA$.  This implies that $\mcA$ is totally geodesic both for $q$
and $\tilde q$.

Choose any point $p$ on $\mcA$. By a shift of $(v,w)$ we can arrange
to have $(v(p),w(p))=(0,0)$. Let $(\rho,z)$ be coordinates obtained
by a rigid rotation of $(v,w)$ around the origin so that the vector
tangent to $\mcA$ at $p$ coincides with $\partial_z$. Then the axis
$\{(0,z)\}_{z\in \R}$ is  a geodesic of $\tilde q$, sharing a common
direction at $p$ with $\mcA$, hence
$$
 \mcA\equiv\{(0,z)\}_{z\in \R}\;.
$$
Since $\phi$ is an isometry of $\tilde q$ which is the identity on
$\mcA$, it easily follows that
$$
 \phi(\rho,z)=(-\rho,z)
 \;,
$$
so that $\MUone=\{\rho\ge 0\}$. We have thus obtained the
representation \eq{axmet2} of $g$.

The reader might have noticed that the function $u$ constructed in
this section is a solution of a Neumann problem with vanishing data
on the axis.

For further use, we note that from \eq{axmet2}, on
$\exp((T\mcA)^\perp)$ the geodesic distance $\hrho$ from the origin
equals
$$
 \hrho =e^{-(U-\alpha)(0,z)}\rho+O(\rho^3)
 \;,
$$
and comparing with \eq{gphirel} we obtain
\bel{alphregcond}
 \alpha(0,z)=0 \;.
\ee
Now, the function $u=o_{k-4}(r^{-1/2})$ of Proposition \ref{Pisocor}
equals $u=2(\alpha-U)$ (compare \eq{axmet2}). By \eq{alphregcond}
and an analysis of Taylor expansions as in Section~\ref{ssAf} we
infer that, at $\{y=0\}$,
\bel{alphregcond2}
 \alpha =o_{k-5}(r^{-3/2})x\;.
\ee
From Proposition~\ref{PAxsymaf} we conclude:

\begin{Theorem}
 \label{TAxsmaf}
Let $k\ge 5$. Any Riemannian metric on $\R^3$ invariant under
rotations around a coordinate axis and satisfying
\bel{decrateg}
  g_{ij}-\delta_{ij}=o_k(r^{-1/2})
\ee
 admits a global representation of the form
\eq{axmet2}, with the functions $U$, $\alpha$, $B_\rho$ and $A_z$
satisfying
\bel{decayr} A_z=
 o_{k-3}(r^{-3/2})\;; \quad B_\rho  =
 o_{k-3}(r^{-5/2})\;; \quad  U  =
 o_{k-3}(r^{-1/2}) \;; \quad \alpha  =
 o_{k-4}(r^{-1/2})\;.
\ee
 Furthermore \eq{alphregcond2} holds.
\end{Theorem}

\begin{Remark}
The decay rate $o(r^{-1/2})$ in \eq{decrateg} has been tailored to
the requirement of a well-defined ADM mass; the result remains true
with  decay rates $o(r^{-\alpha})$ or $O(r^{-\alpha})$ for any
$\alpha\in (0,1)$, with the decay rate carrying over to the
functions appearing in \eq{axmet2} in the obvious way, as in
\eq{decayr}.
\end{Remark}

\subsubsection{Several asymptotically flat ends}
\label{SSe}

The above considerations generalize to  several asymptotically flat
ends:

\begin{Theorem}
 \label{TAxsmaf2}
Let $k\ge 5$, and consider  a simply connected three-dimensional
Riemannian manifold $(M,g)$ which is the union of a compact set and
of $N$ asymptotically flat ends, and let $\Mext$ denote the first
such end. If $g$ is invariant under an action of $\Uone$, then $g$
admits a global representation of the form \eq{axmet2}, where the
coordinates $(z,\rho)$ cover $(\R\times \R^+)\setminus \{\vec
a_i\}_{i=2}^N$, with the punctures $\vec a_i= (0,\puncti )$ lying on
the $z$-axis, each $\vec a_i$ representing ``a point at infinity" of
the remaining asymptotically flat regions. The functions $U$,
$\alpha$, $B_\rho$ and $A_z$ satisfy \eq{decayr} in $\Mext$.

If we   set
$$
r_i=\sqrt{\rho^2 + (z-\puncti )^2}
 \;,
$$
then we  have the following asymptotic behavior near each of the
punctures
%
\bel{asympPunc}
 U  = 2 \ln r_i +
 o_{k-4}(r_i^{1/2}) \;,\quad \alpha  = 
 o_{k-4}(r_i^{1/2})\;,
\ee
where  $f= o_{\ell}(r_i^{1/2})$ means that $
\partial_{A_1}\ldots\partial_{A_j}f=o_{\ell-j}(r_i^{1/2-j})$ for $0\le j\le \ell$.
Finally, \eq{alphregcond} holds.
\end{Theorem}

\proof As discussed in Section~\ref{sSgc}, $M$ is diffeomorphic to
$\R^3$ minus a finite set of points  and, after perfoming a
diffeomorphism if necessary, the action of the group is that by
rotations around a coordinate axis of $\R^3$. As in the proof of
Theorem~\ref{TAxsmaf} there exists a function $v=o_{k-4}(r^{-1/2})$
so that the metric $ e^{-2v} q$
is flat for $|x|$ large enough in each of the asymptotic regions.
\Eq{utwoeq} is then equivalent to the following equation for $u-v$,
\bel{eqag25}
 \Delta_{e^{-2v} q}(u-v)= -e^{2v}\Big(\frac
{R(q)}2+ \Delta_{  q} v \Big)
 \;,
\ee
where the right-hand-side is compactly supported on $\MUone$. Let
$\Mextone$  be the orbit space associated to the first
asymptotically flat region and let $\psi$ be any smooth strictly
positive function on $\MUone$ which coincides with $|\vec y
 |^{-4}$ in each of the remaining asymptotically flat regions of $\MUone$,
where the $y^A$'s are the manifestly flat coordinates there, with
$\psi$ equal to one in $\Mextone$. Then \eq{eqag25} is equivalent to
\bel{eqag26}
 \Delta_{\psi e^{-2v} q}(u-v)=  -\psi^{-1}e^{2v}\Big(\frac
{R(q)}2+\Delta_{  q} v \Big)
 \;.
\ee
Both the metric $\psi e^{-2v}q$ and the source term extend smoothly
through the origins, say $i^0_j$, $j=1,\ldots,N$, of each of the
local coordinate systems $x^A:=y^A/|\vec y|^2$. Simple connectedness
of the two-dimensional manifold
$$
 \omcN:=\MUone\cup\{i^0_j\}_{j=2}^N
$$
implies that $\omcN\approx \R^2$, so that \eq{eqag26} is an equation
to which Lemma~\ref{LExLapl} applies. We thus obtain a solution, say
$w$, of \eq{eqag26}, and subsequently a solution $v+w$ of
\eq{utwoeq} which tends to a constant in each of the asymptotically
flat regions (possibly different constants in different ends),
except (as will be seen shortly) in $\Mext$ where it diverges
logarithmically. Note that at large distances in each of the
asymptotically flat regions the function $w$ is harmonic with
respect to the Euclidean metric, hence approaches its asymptotic
value as $|y|^{-1}$, with gradient falling-off one order faster.
Similarly $v$ has controlled asymptotics there, as in the proof of
Lemma~\ref{LExLapl}. Integrating \eq{utwoeq} over $\MUone$ one finds
that the coefficient of the logarithmic term is again as in
\eq{mudefR}.

In order to determine that coefficient, we note that since
$\omcN\approx \R^2$ there exists a global orthonormal coframe for
$g$, e.g. obtained by a Gram-Schmidt procedure from a global
trivialization of  $T^*\R^2$. As a starting point for  this
procedure one can, and we will do so, use a holonomic basis $dx^A$
with the coordinate functions $x^A$ equal to the manifestly flat
coordinates in $\Mextone$. Furthermore, after
a rigid rotation of the $y^A$'s if necessary, where the $y^A$'s are
the manifestly flat coordinates for the metric $e^{-2(w+v)}q$ in the
asymptotically flat regions other than $\Mextone$,  we can also
assume that the $dx^A$'s coincide with $d(y^A/|\vec y|^2)$ near each
$i^0_j$. By \eq{Rid} and by what is said in the paragraph  following
that equation we have
$$
\mu= \frac 1 {4\pi} \int_{\MUone} R(q) d\mu_q=
\sum_{j=2}^N\lim_{\epsilon \to 0}\frac 1 {2\pi}
\oint_{C(i^0_j,\epsilon)}\omega^1{}_2
 \;.
$$
where the $C(i^0_j,\epsilon)$'s are circles of radius $\epsilon$
centred at the $i^0_j$'s. Near each $ i^0_j $ the metric $q$ takes
the form $e^{2(v+w)}\delta_{AB}dy^A dy^B= e^{2(v+w)}|\vec
x|^{-4}\delta_{AB}dx^A dx^B$. The co-frame $\ttheta ^A$ is given by
$\ttheta ^A=e^{(v+w)}|\vec x|^{-2} dx^A$, leading to
$$
\omega^{1}{}_2 = \frac 2{|\vec x|^{2}} (x^1 dx^2 - x^2 dx^1)+
o(|\vec x|^{-1/2})dx^A
 \;,
$$
so that
$$
\lim_{\epsilon \to 0} \oint_{C(i^0_j,\epsilon)}\omega^1{}_2 = 4\pi
\;.
$$
We note that we have proved:

\begin{Proposition}
 \label{PGaussB}
Let $q$ be a Riemannian metric on a simply connected two-dimensional
manifold which is the union of a compact set and $N$ ends which are
asymptotically flat in the sense of \eq{twodimAFq}, then
$$
\mu:= \frac 1 {4\pi} \int  R(q) d\mu_q=  2(N-1)
 \;.
$$
\qed
\end{Proposition}

Since $\mu\ne 0$, the function $v+w$ obtained so far needs to be
modified to get rid of the logarithmic divergence. In order to do
that for $j=2,\ldots, N$ we construct functions $u_j$, $q$-harmonic
on $\MUone$, such that, in coordinates $x^A$ which are manifestly
conformally flat in each of the asymptotic regions,
\bel{asymptj}
 u_j = \left\{
         \begin{array}{ll}
          \ln |\vec x| +o(1), & \hbox{in $\Mextone$;} \\
         - \ln  |\vec x| +O(1), & \hbox{in the $j$'th asymptotic region;} \\
          O(1), & \hbox{in the remaining asymptotic regions.}
        \end{array}
      \right.
\ee
This can be done as follows: let $\hat u_j$ be any smooth function
which in local manifestly conformally flat coordinates both near
$i^0_j$ and on $\Mextone$ equals $\ln |\vec x|$,   and which equals
one at large distances in the remaining asymptotically flat regions.
Let $\psi$ be as in \eq{eqag26}, then $\Delta_{\psi e^{-2(v+w)}q}
\hat u_j$ is compactly supported in $\MUone$. Further
\beaa \lefteqn{ \int_{\MUone}\Delta_{\psi e^{-2(v+w)}q} \hat u_j
\,d\mu_{\psi e^{-2(v+w)}q} %
}&&
 \\
 &&
= \int_{\MUone}\Delta_{\psi e^{-2v}q} \hat u_j \,d\mu_{\psi
e^{-2v}q}
 \\
 &&
= \lim_{R\to\infty} \int_{C(0,\rho)} D \ln  |\vec x|\cdot n -
\lim_{\epsilon \to 0} \int _{C(0,\epsilon)} D \ln  |\vec x| \cdot n
 \\
 &&
= 0
 \;.
\eeaa
We can therefore invoke Lemma~\ref{LExLapl} to conclude that there
exists a uniformly bounded function $\hat v$, approaching zero as
one recedes to infinity in $\Mextone$, such that
$$
\Delta_{\psi e^{-2(v+w)}q} \hat v = -\Delta_{\psi e^{-2(v+w)}q} \hat
u_j
 \;.
$$
Subsequently the function $u_j:=\hat u_j+\hat v$ is $q$--harmonic
and satisfies \eq{asymptj}.

The function
$$
 u := v+w +2\sum_{j=2}^N u_j+\alpha\;,
$$
where $\alpha$ is an appropriately chosen constant
(compare~\cite{ChAFT}), defines the desired conformal factor
approaching one as one tends to infinity in $\Mextone$ so that
$e^{-2u}q$ is flat. This conformal factor further compactifies each
of the asymptotic infinities except the first one to a point, so
that $e^{-2u}q$ extends by continuity to a flat complete metric on
the simply connected manifold $\omcN$. By the Hadamard-Cartan
theorem there exists on $\omcN$ a global manifestly flat coordinate
system for $e^{-2u}q$. The    axis of rotation can be made to
coincide with a coordinate axis  as in the proof of
Theorem~\ref{TAxsmaf}. It should be clear that the points at
infinity $i^0_j$ lie on  that axis.

In order to prove \eq{asympPunc}, note that the construction above
gives directly.
$$
 U-\alpha = u = C_i + 2 \ln r_i + o_{k-4}(r_i^{1/2})
 \;,
$$
Next, $U$ can be determined by applying an inversion
\bel{inversion}
 y^A\mapsto
 (\rho,z-a_i)=(x^A)=(y^A/|\vec y|^2)
\ee
to \eq{Brho2},
$$
 \rho^2 e^{-U }= g(\eta,\eta) =
\frac{\rho^2}{(\rho^2+(z-a_i)^2)^2}\Big(1+o_{k-3}((\rho^2+(z-a_i)^2)^{1/4})\Big)
 \;.
$$
Since $\alpha$ vanishes on the axis $(y^1)^2+(y^2)^2=0$ in each of
the asymptotic regions, we conclude that $C_i=0$, and \eq{asympPunc}
follows.
\qed

\section{ADM mass} \label{ssADMm}

Let $m$ be the ADM mass of $g$,
$$
m:= \lim _{R\to\infty}\frac 1 {16\pi}
\int_{S_R}(g_{ij,j}-g_{jj,i})dS_i \;,
$$
where $dS_i = \partial_i\rfloor (dx\wedge dy\wedge dz)$. This has to
be calculated in a coordinate system satisfying \eq{falloff1}.
Typically one takes $S_R$ to be a coordinate sphere $S(R)$ of radius
$R$; however, as is well-known, under \eq{falloff1} $S_R$ can be
taken to be any piecewise differentiable surface homologous to
$S(R)$ such that
\bel{ScondR}
 \inf\{r(p)|p\in S_R\}\to_{R\to\infty}\infty\;.
\ee
We will exploit this freedom in our calculation to follow.

We introduce new coordinates $x$ and $y$ so that $\rho$ and
$\varphi$ in \eq{axmet2} become the usual polar coordinates on
$\R^2$:
$$
 x= \rho \cos \varphi\;, \quad y = \rho \sin \varphi
 \;.
$$
This implies
\beaa
 &
 \rho d\rho = \frac 12 d(\rho^2) = x dx + y dy\;,
 &
 \\
 &
 \rho^2 d\varphi = xdy - y dx\;,
 &
 \\
 &
 \rho^2 d\varphi^2 = dx^2 + dy^2   - d\rho^2
 \;.
 &
\eeaa
Inserting the above in \eq{axmet2} one obtains
\bean g &=&  e^{-2U}  (\underbrace{dx^2 +dy^2
)}_{d\rho^2+\rho^2d\varphi^2} +
 \frac{e^{-2U}(e^{
 2\alpha}-1)}{\rho^2}\underbrace{(xdx+ydy)^2}_{\rho^2d\rho^2}
+  e^{-2U+2\alpha}dz^2
 \\
 &   &
 + 2 e^{-2U}(xdy - y dx) \Big(
B_{\rho} (xdx+ydy) + A_z dz \Big)
 \nonumber
 \\
 &&
 + \mbox{ terms quadratic in $(B_\rho,A_z)$}\, .
 \label{axmet2a}
 \eea
This will satisfy \eq{falloff1} if we assume that
\beal{fo4}
 &
 \displaystyle
 U  \;,\   \frac{(e^{ 2\alpha}-1)x^2}{\rho^2}\;,\
  \frac{(e^{ 2\alpha}-1)xy}{\rho^2}\;,\
  \frac{(e^{ 2\alpha}-1)y^2}{\rho^2}\; =
 \;o_1(r^{-1/2})\;,
 &
 \\
 &
 B_\rho x^2\;,\ B_\rho xy\;,\ B_\rho y^2\;,\  A_z x\;,\ A_z y
 \;=\;
 o_1(r^{-1/2})\;,
 &
\eeal{fo5}
consistently with Theorem~\ref{TAxsmaf}. Then the term occurring in
the last line of \eq{axmet2a} will not give any contribution to the
mass integral:
\bean
 g &=& \underbrace{e^{-2U} \left(dx^2 +dy^2 \right)}_{(a)}+
\underbrace{\frac{ e^{ 2\alpha}-1}{\rho^2}(xdx+ydy)^2}_{(b)} +
\underbrace{e^{-2U+2\alpha}dz^2}_{(c)}
 \\
 &&
 +\underbrace{2 (xdy - y dx) \Big(
B_{\rho} (xdx+ydy) + A_z dz \Big)}_{(d)}
 \nonumber
 \\
 &&
 + o_1(r^{-1})dx^idx^j\, .
 \label{axmet2b}
 \eea
Let us denote by $x^a$ the variables $x,y$. As an example, consider
the contribution of   $(c)$ to the mass integrand:
$$
(c)\ \longrightarrow\  g_{zz,z}dS_z-g_{zz,i}dS_i =
-g_{zz,a}dS_a=\Big(2(U-\alpha)_{,a}+o(r^{-2})\Big) dS_a
 \;.
$$
A similar calculation of  $(a)$ easily leads to
$$
(a)+(c)\ \longrightarrow\  (4U_{,i}+ o(r^{-2}))dS_i-2\alpha_{,a}
dS_a
 \;.
$$
The contribution of $(b)$ to the mass integrand looks rather
uninviting at first sight:
\beaa (b) & \longrightarrow & \Big[ \Big(\frac{ e^{
2\alpha}-1}{\rho^2}\Big)_{,y} xy -
 \Big(\frac{ e^{
2\alpha}-1}{\rho^2}\Big)_{,x} y^2 + \frac{ e^{ 2\alpha}-1}{\rho^2} x
\Big] dS_x
 \\
 && +
 \Big[ \Big(\frac{ e^{
2\alpha}-1}{\rho^2}\Big)_{,x} xy -
 \Big(\frac{ e^{
2\alpha}-1}{\rho^2}\Big)_{,y} x^2 + \frac{ e^{
2\alpha}-1}{\rho^2}  y \Big] dS_y \\
 &&
 -\Big(\frac{ e^{
2\alpha}-1}{\rho^2}\Big)_{,z} (x^2+y^2)dS_z \;.
  \eeaa
Fortunately, things simplify nicely if $S_R$ is chosen to be the
boundary of the solid cylinder
\bel{cylinder}
 C_R:=\{-R\le z \le R\;,\ 0\le \rho \le R\}
 \;.
\ee
Then $S_R$ is the union of the bottom $B_R=\{z=-R\;,\ 0\le
\rho\le R\}$, the lid $L_R=\{z=R\;,\ 0\le \rho\le R\}$, and the wall
$W_R=\{-R\le z\le R\;,\ \rho=R\}$. On the bottom and on the lid we
only have a contribution from $dS_z$, which equals
$$
-\Big(2\alpha_{,z} + o(r^{-2})\Big) dx\wedge dy
$$
on the lid, and minus this expression on the bottom. On the wall
$dS_z$ gives no contribution, while
$$ dS_x|_{W_R}= (dy\wedge dz)|_{W_R}= x|_{W_R} d\varphi \wedge
dz\;, \quad dS_y|_{W_R}= -(dx\wedge dz)|_{W_R}= y|_{W_R} d\varphi
\wedge dz
 \;.
$$
Surprisingly, the terms in $(b)|_{W_R}$ containing derivatives of
$\alpha$ drop out, leading to
$$
(b)|_{W_R} \ \longrightarrow \Big(2\alpha  + o(r^{-2})\Big) d
\varphi\wedge dz
$$
We continue with the contribution of $B_\rho$ to   $(d)$:
%
$$
\Big[\underbrace{\Big((x^2-y^2)B_\rho\Big){}_{,y}}_{(1)}
 -
\underbrace{(2xy B_\rho)_{,x}}_{(2)}
 \Big]dS_x + \Big[
\underbrace{\Big((x^2-y^2)B_\rho\Big){}_{,x}}_{(3)}
 +
 \underbrace{(2xy
B_\rho)_{,y}}_{(4)}\Big]dS_y \;.
$$
It only contributes on the wall $W_R$, giving however a zero
contribution there:
\beaa
 \lefteqn{ \Big[ \Big(\underbrace{(x^2-y^2)(x\partial_y+y\partial_x)}_{(1)+(3)}
  +
  \underbrace{2xy
(y\partial_y-x\partial_x)}_{(4)+(2)}\Big)B_\rho \Big] d\varphi
\wedge dz } &&
 \\
 &&
  \phantom{xxxxxxxxxx}=\Big[  (x^2+y^2)\underbrace{(x\partial_y-y\partial_x) B_\rho}_{=0} \Big] d\varphi \wedge
  dz = 0
 \;.
\eeaa
Finally, $A_z$ produces the following boundary integrand:
$$ - y\partial_z A_z dS_x + x\partial_z A_z dS_y +
\Big[\underbrace{(x\partial_y-y\partial_x)A_z}_{=0}\Big]dS_z
 \;,
$$
and one easily checks that the $dS_x$ and $dS_y$ terms cancel out
when integrated upon $W_R$, while giving no contribution on the
bottom and the lid.

Collecting all this we obtain
\beaa m & = & \lim_{R\to\infty} \frac 1 {16 \pi} \Big[ 4\int_{S_R}
\partial_iU dS_i + 2\int_{W_R}(
\alpha -\frac{x^a}{\rho}\partial_a \alpha )\,d\varphi\, dz
 \\
 &&
 - 2\int_{L_R} \partial_z \alpha \,dx\, dy
  +2\int_{B_R} \partial_z \alpha \,dx\, dy
  \Big]
  \\
  & = & \lim_{R\to\infty} \frac 1 {4 \pi} \Big[ \int_{S_R}
\partial_i(U -\frac 12 \alpha)dS_i + \frac 12 \int_{W_R}
\alpha  \,d\varphi\, dz
  \Big]
 \;.
\eeaa
We have the following formula for the Ricci scalar $^{(3)}R$ of the
metric \eq{axmet2}
 (the details of the calculation can be found in
\cite{GibbonsHolzegel}):\footnote{In the time-symmetric case
\eq{diffform} can be viewed as a PDE for $U$ given the remaining
functions and the matter density. Assuming that this equation can
indeed be solved, this allows us to prescribe freely the functions
$\alpha$, $B_\rho$ and $A_z$. In such a rough analysis there does
not seem to be any constraints on $\alpha$, $B_\rho$ and $A_z$ (in
particular they can be chosen to satisfy \eq{fo4}-\eq{fo5}), while
$U$, and hence its asymptotic behavior, is determined by
\eq{diffform}.}
\begin{equation} \label{diffform}
-\frac{e^{-2U+2\alpha}}{4}\; \phantom{}^{(3)}R   = -
 \Delta_\delta  (U -\frac{1}{2} \alpha) + \frac{1}{2} \left( D U \right)^2 - \frac{1}{2\rho}\frac{\partial
\alpha}{\partial \rho} + \frac{\rho^2 e^{-2\alpha}}{8}  \left(\rho
B_{\rho, z} -
 A_{z,\rho}\right)^2 \, .
\end{equation}
The Laplacian  $\Delta_\delta$ and the gradient $D$ are taken with
respect to the flat metric $\delta$ on $\mathbb{R}^3$.

Now,
\bean
 \lefteqn{  \lim_{R\to\infty} \frac 1 {4 \pi} \Big[ \int_{S_R}
\partial_i(U -\frac 12 \alpha)dS_i + \frac 12 \int_{W_R}
\alpha  \,d\varphi\, dz
  \Big]}&&
\\ && =
 \lim_{R\to \infty}\Big[\frac 1 {4\pi}\int_{C_R}
  [\Delta_\delta (U -\frac \alpha 2)+ \frac 1{2\rho} \frac{\partial\alpha}{\partial \rho}] \;
 d^3x
 +\frac 14  \int_{-R}^R \alpha(\rho=0,z)dz\Big]\;.
 \nonumber
 \\
 &&
 \eeal{maxsym2}
The last integral vanishes by \eq{alphregcond}. Equations
\eq{diffform}-\eq{maxsym2} and the dominated convergence theorem
yield now
\bean m&=& \frac{1}{16 \pi} \int \Big[\phantom{}^{(3)}R +
\frac{1}{2} \rho^2
  e^{-4\alpha+2U}\left(\rho B_{\rho, z} - A_{z,\rho}\right)^2 \Big] e^{2(\alpha -U)} d^3 x
 \\
 & & +
\frac{1}{8\pi}\int \left(D U\right)^2 d^3x \, .
 \label{mf}
\eea
Since $\phantom{}^{(3)}R=16 \pi \mu + K_{ab}K^{ab} \geq 0$ for
initial data sets satisfying $\tr_g K=0$, this proves positivity of
mass for  initial data sets as considered above.

Suppose that $m=0$ with $\phantom{}^{(3)}R\ge 0$, then \eq{mf} gives
\bel{vanmasseq}
 \phantom{}^{(3)}R =\rho B_{\rho, z} - A_{z,\rho}=D
U=0
 \;.
\ee
The last equality implies $U\equiv 0$, and from \eq{diffform} we
conclude that
$$
 \Delta_\delta \alpha  -\frac 1 {2\rho} \frac{\partial \alpha}{\partial
 \rho}=0
 \;.
 $$
The maximum principle applied on the set
$$
B(R)\setminus \{\rho \le 1/R\}
$$
gives $\alpha\equiv 0$ after passing to the limit   $R\to \infty$.
The before-last equality in \eq{vanmasseq} shows that the form $\rho
B_\rho d\rho+A_z dz$ is closed, and simple-connectedness implies
that there exists a function $\lambda$ such that $\rho B_\rho
d\rho+A_z dz=d\lambda$, bringing the metric \eq{axmet2} to the form

\begin{equation} \label{axmet2nom}
 d\rho^2 + dz^2  + \rho^2  \left(d(\varphi + \lambda)\right)^2 \, .
\end{equation}
Hence $g$ is flat. One could now attempt to analyse
$\varphi+\lambda$ near the axis of rotation to conclude that
$(\rho,\varphi+\lambda,z)$ provide a new global polar coordinate
system, and deduce that $g$ is the Euclidean metric. However, it is
simpler to invoke the Hadamard-Cartan theorem to achieve that
conclusion.

\bigskip

Summarizing, we have proved:

\begin{Theorem}
\label{Taxmass} Consider a  metric of the form \eq{axmet2} on
$M=\R^3$, where $(\rho,\varphi,z)$ are polar coordinates, with
Killing vector $\partial_\varphi$, and suppose that the decay
conditions \eq{fo4}-\eq{fo5} hold. If
$$
  {}^3R\ge 0
 $$
then $ 0\le m\le \infty$. Furthermore, we have $m<\infty $ if and
only if
$$
{}^3R \in L^1(\R^3)\;, \qquad DU,\,\rho B_{\rho, z} - A_{z,\rho} \in
L^2(\R^3)\;.
 $$
Finally, $m=0$   if and only if $g$ is the
 Euclidean metric.\qed
\end{Theorem}
\begin{Remark}
\label{Rtpe} Theorem \ref{TAxsmaf} shows that the coordinates
required above will exist for a general asymptotically flat
axisymmetric metric on $\R^3$ if \eq{falloff1} holds with $k=6$.
\end{Remark}

 \subsection{Several asymptotically flat ends}

Theorem~\ref{Taxmass} proves positivity of mass for axi-symmetric
metrics on $\R^3$. More generally, one has the following:

\begin{Theorem}
\label{Taxmassmany}
  Let $(M,g)$ be a simply connected three dimensional Riemannian manifold
which is the union of a compact set and of  a finite number of
asymptotic regions $M_i$, $i=1,\ldots,N$, which are asymptotically
flat in the sense of \eq{falloff1}-\eq{falloff1a} with $k\ge 6$. If
$g$ is invariant under an action of $\Uone$ and
$$
  {}^3R\ge 0\;,
 $$
then the ADM mass $m_i$ of each of the ends $M_i$ satisfies $ 0<
m_i\le \infty$, with $m_i<\infty $ if and only if
$$
{}^3R \in L^1(M_i)\;, \qquad DU,\,\rho B_{\rho, z} - A_{z,\rho} \in
L^2(M_i)\;.
 $$
\end{Theorem}

\proof The result follows immediately from the calculations in this
section together with Theorem \ref{TAxsmaf2}: Indeed, one can
integrate \eq{diffform} on a set
$$
 \hat C_R:= C_R \setminus C_{1/R} =\{-R\le z \le R\;,\ 1/R\le \rho
 \le R\}
 \;,
$$
where $C_R$ is as in \eq{cylinder}. The asymptotics \eq{asympPunc}
implies that the boundary integrals over the boundary of $C_{1/R}$
give zero contribution in the limit $R\to \infty$, so that  \eq{mf}
remains valid  by the monotone convergence theorem in spite of the
(mildly) singular behavior at the punctures $\vec a_i$ of the
functions appearing in the metric. \qed

\subsection{Nondegenerate \ih s}
\label{Smashor}

In order to motivate the boundary conditions in this section, recall
that in Weyl coordinates the Schwarzschild metric takes the form
(\cf, \eg~\cite[Equation~(20.12)]{Exactsolutions2})
 \bel{SchwWeyl}
\fourg = -e^{2\uS} dt^2 + e^{-2\uS} \rho^2 d\varphi^2 +
e^{2\lS}(d\rho^2 + dz^2)
 \;,
\ee
where
%
\beal{SchwuW}
 \uS &=&\ln \rho -\ln\Big( m\sin \ttheta  +\sqrt{\rho^2+m^2\sin^2\ttheta
 }\Big)
 \\
 \label{SchwuNM}
 & =&
   \frac 12 \ln \left[ \frac{\sqrt{(z-m)^2+\rho^2}
 +\sqrt{(z+m)^2+\rho^2} - 2m}{\sqrt{(z-m)^2+\rho^2}
 +\sqrt{(z+m)^2+\rho^2} + 2m} \right]
\;,
 \\
 \lS
  & = & - \frac 12
  \ln \left[ \frac{(\rS-m)^2-m^2 \cos^2\ttheta}{\rS^2}\right]
 \\
  & = & - \frac 12
  \ln \left[ \frac{4\sqrt{(z-m)^2+\rho^2}
  \sqrt{(z+m)^2+\rho^2}}
  {\Big[2m+\sqrt{(z-m)^2+\rho^2}
 +\sqrt{(z+m)^2+\rho^2}\Big]^2}\right]
  \;.
 \eea
In \eq{SchwuW} the angle $\ttheta $ is a  Schwarzschild angular
variable, with the relations
\beaa
 &
 2m \cos \ttheta  =  \sqrt{(z+m)^2+\rho^2}
 -\sqrt{(z-m)^2+\rho^2}
 \;,
 &
\\
 &
 2(\rS-m)=  \sqrt{(z+m)^2+\rho^2}
 +\sqrt{(z-m)^2+\rho^2}
 \;,
&
 \\
 &
\rho^2 = \rS(\rS-2m)\sin^2\ttheta \;, \quad z=(\rS-m)\cos \ttheta
 \;,
 &
\eeaa
where $\rS$ is the usual Schwarzschild radial variable such that
$e^{2\uS}=1-2m/\rS$. As is well known, and in any case easily seen,
$\uS$ is smooth on $\R^3$ except on the set $\{\rho=0, -m\le z \le
m\}$. From \eq{SchwuW} we find, at fixed $z$ in the interval $-m < z
< m$ and for small $\rho$,
\bel{InstHor0}
 \uS(\rho,z) = \ln \rho -\ln (2\sqrt{(m+z)(m-z)})+ O(\rho^2)
\ee
%
(with the error term \emph{not }uniform in $z$). This justifies our
definition: an interval $[a,b]\subset \mcA$ will be said to be a
\emph{\nih}\ if for fixed $z\in (a,b)$ and for small $\rho$ we have
\bel{InstHor}
  U(\rho,z)= \ln \rho +\zU(z)+ o(1)\;, \quad \partial U(\rho,z) = \partial  \ln \rho +\partial \zU(z)+ o(1)\;,
\ee
%
for a smooth function $\zU$.  As in the Schwarzschild case the
function $U-\alpha$ is assumed to be smooth across $I$. Thus, to
compensate for the logarithmic singularity of $U$, we further
assume, again for fixed $z\in (a,b)$ and for small $\rho$, that
there exists a function $\zlambda(z)$ such that
\bel{InstHora}
 \alpha(\rho,z) = U(\rho,z) + \zlambda(z) + o(1)
\;. \ee
%
Under those conditions the calculation of the mass formula proceeds
as follows. For $k=1, \ldots, N$ let
$$
 I_k = [c_k,d_k] \subset \mcA
$$
be pairwise disjoint intervals at which the \nih\ boundary
conditions hold.  Denote by $\tU$ the function, harmonic on $\R^3
\setminus \cup_k I_k$, which is the sum of Schwarzschild potentials
$\uS$ as in \eq{SchwuNM}, each with mass $(d_k-c_k)/2$ and a
logarithmic singularity at $I_k$. As in~\cite{GibbonsHolzegel}, the
term $|DU|^2$ in \eq{diffform} is rewritten as:
$$
|DU|^2= |D(U-\tU+\tU)|^2 = |D(U-\tU )|^2 + D_i\Big[
 (2U-\tU)D^i\tU\Big]
\;.
$$
Denote by $I_\epsilon$ the set of points which lie a distance less
than or equal to $\epsilon$ from the singular set $\cup_k I_k$:
$$
 I_\epsilon = \{p\ |\ d(p,\cup_kI_k)\le \epsilon\}
 \;.
$$
By inspection of the calculations  so far one finds that \eq{mf}
becomes now
\bean m&=& \frac{1}{16 \pi} \int \Big[\phantom{}^{(3)}R +
\frac{1}{2} \rho^2
  e^{-4\alpha+2U}\left(\rho B_{\rho, z} - A_{z,\rho}\right)^2 \Big] e^{2(\alpha -U)} d^3 x
 \\
 & & +
\frac{1}{8\pi}\int \left(D (U-\tU)\right)^2 d^3x
 \nonumber
 \\
 &&
+ \frac{1}{8\pi} \lim_{\epsilon\to 0} \int_{\partial
I_\epsilon}\Big[ D^i(2U-\alpha)-(2U-\tU)D^i \tU +   \alpha \frac
{D^i\rho}\rho\Big]n_id^2S\, .
 \label{mf2.0}
\eea
In the last line of \eq{mf2.0} the normal $n_i$, taken with respect
to the flat metric, has been chosen to point away from $I_\epsilon$.

Away from the end points of the intervals $I_k$ the logarithmic
terms in $U$, $\tU$ and $\alpha$ cancel out, leaving a contribution
$$
\frac{1}{4}  \sum_k\left(|I_k|+\int_{I_k}( \zlambda+\zbeta
)dz\right)\, ,
$$
%
where $|I_k|$ is the length of $I_k$, and where we have denoted by
$\zbeta$ the limit at $\cup_k I_k$ of $\tU-U$,
%
$$
 \zbeta(z):= \lim_{\rho\to 0\;,\ z\in \cup_kI_k} \left(\tU(\rho,z)-U(\rho,z)\right)
 \;.
$$
As already pointed out, the error term in \eq{InstHor0} is not
uniform in $z$, and therefore it is not clear whether or not there
will be a separate contribution from the end points of $I_k$ to the
limit as $\epsilon$ tends to zero of the integral over $\partial
I_\epsilon$. Assuming that no such contribution arises%
\footnote{Note that this assumption, asymptotic flatness, finiteness
of the volume integral in \eq{mf2.0}, and the boundary condition
\eq{InstHor} on $U$ essentially enforce the boundary condition
\eq{InstHora} on $\alpha$.},
we conclude that the following formula for the mass holds:
\bean m&=& \frac{1}{16 \pi} \int \Big[\phantom{}^{(3)}R +
\frac{1}{2} \rho^2
  e^{-4\alpha+2U}\left(\rho B_{\rho, z} - A_{z,\rho}\right)^2 \Big] e^{2(\alpha -U)} d^3 x
 \\
 & & +
\frac{1}{8\pi}\int \left(D (U-\tU)\right)^2 d^3x
 \nonumber
\\
 &&
 +\frac 14 \sum_k\left(|I_k|+\int_{I_k}( \zlambda+\zbeta
)dz\right) \, .
 \label{mf3}
\eea
%
In the Schwarzschild case the volume integrals vanish, $\zbeta=0$,
for  $z\in(-m,m)$ the function $\zlambda$ equals
$$ \zlambda (z)
  =
   - \frac 12
  \ln \left[ \frac{(m-z)(z+m) }
  {(2m)^2}\right]
  \;,
$$
and one can check \eq{mf3} by a direct calculation of the integral
over $I_1$.

\subsection{Conical singularities}
 \label{Srods}


So far we have assumed that the metric is smooth across the rotation
axis $\mcA$. However, in some situations this might not be the case.
One of the simplest examples is the occurrence of conical
singularities, when the regularity condition \eq{alphregcond}  fails
to hold. It is not clear what happens with the construction of the
coordinates~\eq{axmet2} in such a case, and therefore it appears
difficult to make general statements concerning such metrics.
Nevertheless, there is at least one instance where conical
singularities occur naturally, namely in the usual construction of
stationary axisymmetric solutions: here one assumes at the outset
that the space-time metric takes a form which reduces to \eq{axmet2}
after restriction to slices of constant time; and the components of
the metric are then obtained by various integrations starting from a
solution of a harmonic map equation; \cf,
\eg~\cite{Weinstein2,Neugebauer:2003qe,Ernst}.

So consider a metric of the form~\eq{axmet2} on $\R^3\setminus\{\vec
a_i\}$, where each puncture $\vec a_i$ corresponds to either an
asymptotically flat region  or to asymptotically cylindrical regions
 (which, typically, correspond to degenerate black holes).
Assuming that $d\alpha$ is bounded at the axis and does not give any
supplementary contribution at the punctures, \eq{mf} becomes instead
\bean m&=& \frac{1}{16 \pi} \int_{\R^3\setminus\{\vec a_i\}}
\Big[\phantom{}^{(3)}R + \frac{1}{2} \rho^2
  e^{-4\alpha+2U}\left(\rho B_{\rho, z} - A_{z,\rho}\right)^2 \Big] e^{2(\alpha -U)} d^3 x
 \\
 & & +
\frac{1}{8\pi}\int_{\R^3\setminus\{\vec a_i\}} \left(D U\right)^2
d^3x
 + \frac 14 \int_{\mcA\setminus\{\vec a_i\}} \mathring \alpha \;dz
\, ,
 \label{mf2s}
\eea
where $\mathring \alpha$ denotes the restriction of $\alpha$ to
$\mcA$.

Using \eq{mf2s} and \eq{mf3}, the reader will easily work out a mass
formula when both conical singularities and \nih s occur.

\bigskip

\noindent{\sc Acknowledgements:} I thank Sergio Dain and Gustav
Holzegel for useful discussions. I am grateful to the AEI for
hospitality and support during the early stage of work on this
paper.

\bibliographystyle{amsplain}
\bibliography{../references/hip_bib,%
../references/reffile,%
../references/newbiblio,%
../references/newbiblio2,%
../references/bibl,%
../references/howard,%
../references/bartnik,%
../references/myGR,%
../references/newbib,%
../references/Energy,%
../references/netbiblio}
\end{document}